\documentclass[preprint,aps,amsmath,showpacs,floatfix]{revtex4}
\newcommand{\rd}{{\rm d}}
\newcommand{\re}{{\rm e}}
\newcommand{\ri}{i}

\usepackage{graphicx}

\begin{document}

\title{Coherent State Path Integrals in the Weyl Representation}

\author{L.C. dos Santos and M.A.M. de Aguiar }

\affiliation{Instituto de F\'{\i}sica `Gleb Wataghin'\\
Universidade Estadual de Campinas, Unicamp\\
13083-970, Campinas, S\~{a}o Paulo, Brasil\\}

\begin{abstract}

We construct a representation of the coherent state path integral
using the Weyl symbol of the Hamiltonian operator. This
representation is very different from the usual path integral forms
suggested by Klauder and Skagerstan in \cite{Klau85}, which involve
the normal or the antinormal ordering of the Hamiltonian. These
different representations, although equivalent quantum mechanically,
lead to different semiclassical limits. We show that the
semiclassical limit of the coherent state propagator in Weyl
representation is involves classical trajectories that are
independent on the coherent states width. This propagator is also
free from the phase corrections found in \cite{Bar01} for the two
Klauder forms and provides an explicit connection between the Wigner
and the Husimi representations of the evolution operator.

\end{abstract}



\maketitle

\section{Introduction}

The set of coherent states forms a non-orthogonal over-complete
basis. This has important consequences for the path integral
formulation of the propagator. It implies the existence of several
forms of path integrals, all equivalent quantum mechanically, but
each leading to a slightly different semiclassical limit. Klauder
and Skagerstam (KS) \cite{Klau85} proposed two basic forms for the
coherent state path integral, whose semiclassical limits were
considered in \cite{Bar01}. It was shown in \cite{Bar01} that these
two semiclassical propagators can written in terms of classical
complex trajectories, each governed by a different classical
representation of the Hamiltonian operator $\hat{H}$: the P
representation $H_P$ in one case and the Q representation $H_Q$ in
the other. We briefly review these representations and their
semiclassical limits in section 2. The two most important
characteristics of these semiclassical formulas are, first, that the
underlying classical dynamics depends explicitly on the width of
coherent states. Second, the phase appearing in these semiclassical
formulas is not just the action of the corresponding complex
classical trajectory, but it also contains a `correction term' $I$
that comes with different signs in each formula (see
Eqs.(\ref{glg119}) and (\ref{glg119a})).

In \cite{Bar01} it was also suggested that a semiclassical
representation involving directly the Weyl representation of
$\hat{H}$, or the {\it classical Hamiltonian} $H_W$, could probably
be constructed, and a formula for this representation was
conjectured. A first attempt to derive such formula was recently
presented in \cite{coelho}. The strategy used there was to build the
propagator out of infinitesimal propagators that alternated between
the two KS forms. The resulting semiclassical dynamics turned out to
be governed by $(H_Q+H_P)/2$, which coincides with the Weyl symbol
for polynomial Hamiltonians with up to cubic terms in $q$ and $p$
only. The correction to the action was found to be $(I_Q-I_P)/2$,
which is also non-zero for general Hamiltonians. In this paper we
construct a new representation of the quantum mechanical path
integral in the coherent state representation that contains
precisely $H_W$ and derive its semiclassical limit. The new
construction is based on the properties of translation and
reflection operators \cite{balazs,Alf98}, which form basis for
expressing general operators. While in the KS path integrals each
path contributes a term of the form $\exp{iS/\hbar}$, where $S$ is
the action along the path (computed with either $H_Q$ or $H_P$), the
exponent in the new form is rather different and does not
immediately resembles an action. Although the terms in this exponent
can be re-arranged so as to look similar to the action function, it
is only when the limit of continuous paths is taken that one can
really recognize the action as a part of the exponent.

We show that the semiclassical limit of the coherent state
propagator in the Weyl representation is indeed given by the
expression conjectured in \cite{Bar01}: the underlying dynamics is
purely classical (independent on the width of the coherent states)
and there is no correction term to be added to the action. More
importantly, the new path integral representation allows for a
direct connection between the coherent state representation of the
evolution operator and its Weyl symbol.

The paper is organized as follows: in section 2 we review the path
integral constructions of Klauder and Skagerstan and their
semiclassical approximations. In section 3 we construct the new path
integral representation and in section IV we derive its
semiclassical limit. The two path integrals of Klauder and
Skagerstan are compared with the new form in section V, where we
also comment on relevance of these results for numerical
calculations. Finally, in section VI, we discuss the connection
between the Weyl symbol of the evolution operator and the diagonal
coherent state propagator.

\section{The coherent state propagator and its semiclassical
approximations}

In this section we define the coherent state propagator and review
the construction of the two path integrals suggested by Klauder and
Skagerstan, showing how the symbols $H_Q$ and $H_P$ of the operator
$\hat{H}$ appear in each of them. We also write down the
semiclassical limit of these path integrals to compare with our
results in the next section. Our presentation here is strongly based
in \cite{coelho}.

\subsection{The propagator}
The coherent state $|z\rangle$ of a harmonic oscillator of mass $m$ and
frequency $\omega$ is defined by
\begin{equation}
  \label{glg48}
  |z\rangle = \re^{-\frac{1}{2}|z|^2}\re^{z\hat{a}^\dagger}|0\rangle
\end{equation}
with $|0\rangle$ the harmonic oscillator ground state and
\begin{equation}
  \label{glg49}
  \hat{a}^\dagger = \frac{1}{\sqrt{2}}\left( \frac{\hat{q}}{b}-\ri
              \,\frac{\hat{p}}{c} \right), \qquad
  z =  \frac{1}{\sqrt{2}}\left( \frac{q}{b}+\ri
              \,\frac{p}{c} \right).
\end{equation}
In the above $\hat q$, $\hat p$, and $\hat{a}^\dagger$ are
operators; $q$ and $p$ are real numbers; $z$ is complex.  The
parameters $b = {(\hbar/ m \omega )}^{\frac{1}{2}}$ and $c =
{(\hbar m \omega )}^{\frac{1}{2}}$ define the length and momentum
scales, respectively, and their product is $\hbar$.

For a time-independent Hamiltonian operator $\hat{H}$, the propagator
in the coherent states representation is the matrix element of the
evolution operator between the states $|z^\prime \rangle$ and $| z''
\rangle$:
\begin{equation}
  \label{glg51}
  K(z'',z',T) = \langle z'' | \re^{-\frac{\ri }{\hbar}\hat{H}T}
                  | z^\prime \rangle.
\end{equation}
We restrict ourselves to Hamiltonians that can be expanded in a
power series of  the creation and annihilator operators
$\hat{a}^\dagger$ and $\hat{a}$.

In the construction of a path integral for $K$, and also in the
derivation the semiclassical limit of the propagator, the Hamiltonian
operator $\hat{H}$ is somehow replaced by a classical Hamiltonian
function $H(q,p)$. This `replacement', however, is not uniquely
defined, and the ambiguities that exist in the relation between the
operator $\hat{H}$ and the function $H(q,p)$ also arise in connection
with the overcompleteness of the coherent state basis, as we shall see
in the next subsections.

There are actually many ways to associate a classical function of
position and momentum $A(q,p)$ to a quantum mechanical operator
$\hat{A}$ \cite{grosche}. However, three of them are specially
important. The first one, denoted $A_Q(q,p)$ and called the Q
representation of the operator $\hat{A}$, is constructed as follows:
one writes $\hat{A}$ in terms of the creation and annihilation
operators $\hat{a}^\dagger$ and $\hat{a}$ in such a way that all the
creation operators appear to the left of the annihilation operators,
making each monomial of $\hat{A}$ look like $c_{nm}\hat{a}^{\dagger
n} {\hat{a}}^m$. Then we replace $\hat{a}$ by $z$ and
$\hat{a}^\dagger$ by $z^\star$. The inverse of this operation, that
associates a quantum operator to a classical function, is called
`normal ordering'. In this case one first writes the classical
function in terms of $z$ and $z^\star$, with all the $z^\star\,$'s
to the left of the $z$'s, and then replace $z$ by $\hat{a}$ and
$z^\star$ by $\hat{a}^\dagger$.

The second possibility, called the P representation of $\hat{A}$,
is obtained by a similar procedure, but this time the monomials of
$\hat{A}$ are written in the opposite order, such that they look
like $c_{nm}\hat{a}^n \hat{a}^{\dagger m}$. Once the operator has
been put in this form one replaces again $\hat{a}$ by $z$ and
$\hat{a}^\dagger$ by $z^\star$ to obtain $A_P(q,p)$. The inverse
of this operation is called `anti-normal ordering'. Notice that
the differences between the two representations come from the
commutator of $\hat{q}$ and $\hat{p}$, which is proportional to
$\hbar$. Therefore, these differences go to zero as $\hbar$ goes
to zero.

There is, finally, a third representation which is the most
symmetric of all, and therefore the most natural. It is given by
the Wigner transformation
\begin{equation}
  \label{wig1}
A_W(q,p) = \int\rd s\,\re^{\frac{\ri}{\hbar}ps}\left\langle
q-\frac{s}{2}\left|
   \hat{A}\right|q+\frac{s}{2}\right\rangle \;.
\end{equation}
$A_W(q,p)$ is called the Weyl representation of $\hat{A}$
\cite{Hill84,Alf98}. Its inverse transformation consists in
writing the classical function in terms of $z$ and $z^\star$
considering all possible orderings for each monomial and making a
symmetric average between all possibilities before replacing $z$
and $z^\star$ by the corresponding operators. As an illustration
of these three representations we take
\begin{displaymath}
\hat{H} = -\frac{1}{2} \frac{\partial^2}{\partial x^2} + \frac{1}{2}
x^2 + x^4
\end{displaymath}
($m=\hbar=1$) for which we obtain
\begin{displaymath}
\begin{array}{l}
H_Q = \frac{1}{2} (p^2 + x^2) + x^4 + \frac{1}{4}(b^2+b^{-2}) + 3b^2x^2 + 3b^4/4 \\
H_P = \frac{1}{2} (p^2 + x^2) + x^4 - \frac{1}{4}(b^2+b^{-2}) - 3b^2x^2 + 3b^4/4 \\
H_W = \frac{1}{2} (p^2 + x^2) + x^4
\end{array}
\end{displaymath}
where $b$ is the width of the coherent state. Notice the term
proportional to $x^2$ that appears with opposite signs in $H_Q$
and $H_P$, really modifying the classical dynamics with respect to
$H_W$.

\subsection{Basic Path Integrals and their Semiclassical Approximations}

The calculation of the semiclassical propagator in the coherent
state representation starting from path integrals was discussed in
detail in \cite{Bar01}. In this section we summarize these
previous results emphasizing the non-uniqueness of the
semiclassical limit as a consequence of the overcompleteness of
the coherent state representation. The reader is referred to
\cite{Bar01} for the details.

In order to write a path integral for $ K(z'',T;z',0)$, the time
interval has to be divided into a large number of slices and, for
each slice, an infinitesimal propagator has to be calculated. As
pointed out by Klauder and Skagerstam \cite{Klau85,Klau78}, there
are at least two different ways to do that. Each of these gives rise
to a different representation of the path integral. Although they
correspond to identical quantum mechanical quantities, their
semiclassical approximations are different. We review the
construction of these two representations below.

The first form of path integral is constructed by breaking the time
interval $T$ into $N$ parts of size $\tau$ and inserting the unit
operator
\begin{equation}
  \label{glg21}
  \openone = \int|z\rangle\frac{\rm{d}z\,\rm{d}z^*}{2\pi i}\langle z|
\end{equation}
everywhere between adjacent propagation steps. We denote the real
and imaginary parts of $z$ by $x$ and $y$, respectively. In all
integrations, ${\rm d}z\,{\rm d} z^*/2\pi i$ means $\rd x\rd y/\pi$.
After the insertions, the propagator becomes a $2(N-1)$--fold
integral over the whole phase space
\begin{equation}
  \label{glg82}
  K(z'',t;z',0) = \int \Bigl\{ \prod_{j=1}^{N-1}\frac{\rm{d}z_j\,\rm{d}z_j^*}{2\pi i}
                   \Bigr\} \prod_{j=0}^{N-1} \Bigl\{ \langle z_{j+1} |
                   \re^{-\frac{\ri }{\hbar}
                   \hat{H}(t_j)\tau} | z_j \rangle
                   \Bigr\}
\end{equation}
with $z_N = z''$ and $z_0=z'$. Using the coherent state overlap
formula
\begin{equation}
  \label{glg50}
 \langle z_{j+1} | z_j \rangle
                        = \exp\left\{ -\frac{1}{2}|z_{j+1}|^2
                        + z_{j+1}^\star z_j
                        - \frac{1}{2}|z_j|^2\right\}
\end{equation}
and expanding $e^{-iH\tau/\hbar} \approx 1 -iH\tau/\hbar$ we write
\begin{equation}
  \label{glg2a}
       \langle z_{j+1} |  \re^{-\frac{\ri }{\hbar}
                   \hat{H}(t_j)\tau} | z_j \rangle
                 = \exp{ \left\{ \frac{1}{2}(
                  z_{j+1}^\star - z_j^\star )z_j - \frac{1}{2}
                  z_{j+1}^\star (z_{j+1} - z_j)-\frac{\ri \tau }
                 {\hbar}{\cal H}_{j+1,j} \right\}}
\end{equation}
where
\begin{equation}
  \label{glg40}
  {\cal H}_{j+1,j} \equiv \frac{\langle z_{j+1}|\hat{H}(t_j)|z_j\rangle }
                   {\langle z_{j+1}|z_j\rangle }
                   \equiv {\cal H}(z_{j+1}^\star,z_j;t_j)
\end{equation}
and $(1 -i {\cal H}_{j+1,j}\tau/\hbar)$ has been approximated again by
$e^{-i {\cal H}_{j+1,j}\tau/\hbar}$.  With these manipulations the
first form of the propagator, that we shall call $K_Q$, becomes
\begin{align}
  \label{glg82a}
  K_Q(z'',t;z',0) = \int \Bigl\{ \prod_{j=1}^{N-1}\frac{\rm{d}z_j\,\rm{d}z_j^*}{2\pi i}
                   \Bigr\}
                  \exp{ \left\{ \sum_{j=0}^{N-1} \left[ \frac{1}{2}(
                  z_{j+1}^\star - z_j^\star )z_j - \frac{1}{2}
                  z_{j+1}^\star (z_{j+1} - z_j)-\frac{\ri \tau }
                 {\hbar}{\cal H}_{j+1,j}\right] \right\}}
\end{align}

When the limit $N \rightarrow \infty$ and $\tau \rightarrow 0$ is
taken, the above summations turn into integrals. Also, ${\cal
H}_{j+1,j}$ turns into the smooth Hamiltonian function ${\cal
H}(z,z^\star) \equiv \langle z |\hat{H}|z\rangle$. Using the
properties $\hat{a}|z\rangle = z|z\rangle$ and $\langle
z|\hat{a}^\dagger = \langle z | z^\star\,$, we see that ${\cal H}$ can
be easily calculated if $\hat{H}$ is written in terms of creation and
annihilation operators with all $\hat{a}^\dagger\,$'s to the left of
the $\hat{a}\,$'s. Therefore, ${\cal H}$ is exactly $H_Q(z,z^\star)$,
the Q symbol of the Hamiltonian operator \cite{Hill84}.

The second form of path integral starts from the ``diagonal
representation'' of the hamiltonian operator, namely
\begin{equation}
  \label{5mb1}
\hat{H} = \int|z\rangle h(z^\star,z)\frac{\rm{d}z\,\rm{d}z^*}{2\pi
i} \langle z| \; .
\end{equation}
Assuming that $\hat{H}$ is either a polynomial in $p$ and $q$ or a
converging sequence of such polynomials, this diagonal
representation always exists. The calculation of $h$ is not as
direct as that of ${\cal H}$, but it can be shown \cite{Hill84} that
$h(z^\star,z)$ is exactly $H_P$, the P symbol of $\hat{H}$. To
facilitate the comparison between this form of path integral, that we
call $K_P$, and $K_Q$, it is convenient to break the time interval $T$
into $N-1$ intervals, rather than $N$. We write
\begin{eqnarray}
 \label{form2a}
K_P(z'',T;z',0)  = \langle z'' | \prod_{j=1}^{N-1}
\re^{-\frac{i\tau}{\hbar}\hat{H}}\,|  z' \rangle
\end{eqnarray}
and, following Klauder and Skagerstam, we write the infinitesimal
propagators as
\begin{equation}
  \label{5mb7}
\re^{-\frac{\ri}{\hbar}\hat{H}\tau} \approx
\int|z_j\rangle
  \left( 1-\frac{i\tau}{\hbar}h(z_j^\star,z_j) \right)
   \frac{\rm{d}z_j\,\rm{d}z_j^*}{2\pi i}\langle z_j|
\approx \int|z_j\rangle
   \re^{-\frac{i\tau}{\hbar}h(z_j^\star,z_j)}
   \frac{\rm{d}z_j\,\rm{d}z_j^*}{2\pi i}\langle z_j| \,.
\end{equation}
The complete propagator $K_P$ becomes
\begin{align}
  \label{5mb8a}
K_P(z_N,&T;z_0,0) =  \displaystyle{
    \int  \prod_{j=1}^{N-1}\frac{\rm{d}z_j\,\rm{d}z_j^*}{2\pi i}
     \langle z_{j+1} | z_j \rangle \,
     \exp{ \left\{ -\frac{\ri \tau } {\hbar} h(z_j^\star,z_j) \right\}}}
     \nonumber \\
 &=  \int  \{ \prod_{j=1}^{N-1}  \frac{\rm{d}z_j\,\rm{d}z_j^*}{2\pi i} \}
             \exp{ \left\{  \sum_{j=0}^{N-1} \left[ \frac{1}{2}(
                  z_{j+1}^\star - z_j^\star )z_j - \frac{1}{2}
                  z_{j+1}^\star (z_{j+1} - z_j)
             -\frac{\ri \tau } {\hbar} h(z_j^\star,z_j) \right]
              \right\}}  \; .
\end{align}

Notice that while the two arguments of $H_Q$ in $K_Q$ belong to two
adjacent times in the mesh, the two arguments of $H_P$ in $K_P$
belong to the same time. Although both forms should give identical
results when computed exactly, the differences between the two are
important for the stationary exponent approximation, resulting in
different semiclassical propagators. The semiclassical evaluation of
$K_Q$ and $K_P$ were presented in detail in \cite{Bar01} (see also
\cite{Klau79,Klau87a,Weis82b}). Here we only list the results:
\begin{eqnarray}
  \label{glg119}
  K_Q(z'',t;z',0) = \sum_\nu
         \sqrt{\frac{\ri}{\hbar}\frac{\partial^2 S_{Q\nu}}
         {\partial u' \partial v''}} \;
         \exp \left\{ \frac{\ri}{\hbar}(S_{Q\nu}+I_{Q\nu}) - \frac{1}{2}
          \bigl( |z''|^2 + |z'|^2\bigr) \right\} \, ,
\end{eqnarray}
\begin{eqnarray}
  \label{glg119a}
  K_P(z'',t;z',0) = \sum_\nu
         \sqrt{\frac{\ri}{\hbar}\frac{\partial^2 S_{P\nu}}
         {\partial  u' \partial v''}} \;
         \exp \left\{ \frac{\ri}{\hbar}(S_{P\nu}-I_{P\nu}) - \frac{1}{2}
          \bigl( |z''|^2 + |z'|^2\bigr) \right\} \, ,
\end{eqnarray}
where
\begin{eqnarray}
  \label{glg83}
  S_{i\nu} = S_{i\nu}(v'',u',t)
         &=\int\limits_0^t \rd t' \left[\frac{\ri \hbar}{2}
          (\dot{u}v-\dot{v}u) - H_i(u,v,t') \right]
          - \frac{\ri \hbar}{2} ( u''v'' + u'v' )
\end{eqnarray}
is the action and
\begin{equation}
\label{epslb}
I_i = \frac{1}{2} \int_0^T
\frac{\partial^2 H_i}{\partial u \partial v}
{\rm d}t
\end{equation}
is a correction to the action. The index $i$ assume the values $Q$
and $P$ and sum over $\nu$ represents the sum over all
`contributing' (complex) classical trajectories satisfying
Hamilton's equations
\begin{eqnarray}
  \label{glg10}
 \ri \hbar\dot{u}&= + \displaystyle{\frac{\partial H_i}{\partial v}}\nonumber\\
 \ri \hbar\dot{v}&= - \displaystyle{\frac{\partial H_i}{\partial u}}
\end{eqnarray}
with boundary conditions
\begin{equation}
\label{mb12} u(0) = z'\equiv u'~, \qquad v(t) = {z''}^\star
\equiv v"~ \;.
\end{equation}
The factors $I_i$ are an important part of the formulas and they are
absolutely necessary to recover the exact propagator for quadratic
Hamiltonians. If one neglects it, even the Harmonic oscillator comes
out wrong. For a discussion about contributing and non-contributing
trajectories, see refs. \cite{Rib04,Ada89}.

Finally we remember that the Weyl Hamiltonian can be obtained from
$\hat{H}$ by completely symmetrizing the creation and annihilation
operators. It turns out to be an exact average between $H_Q$ and
$H_P$ {\it if} $\hat{H}$ contains up to cubic monomials in $\hat{a}$
and $\hat{a}^\dagger$, but only an approximate average for other
cases. The semiclassical formula with $H_Q$ comes with a correction
$+I_Q$ and that with $H_P$ comes with a correction of $-I_P$. This
suggests a third type of semiclassical approximation for the
propagator, where one uses the Weyl Hamiltonian and no correction
term, since the average of $+I_1$ and $-I_2$ should be approximately
zero. This is the Weyl approximation, which was conjectured in
\cite{Bar01}:
\begin{eqnarray}
  \label{glg119b}
  K_W(z'',t;z',0) = \sum_\nu
         \sqrt{\frac{\ri}{\hbar}\frac{\partial^2 S_{W}}
         {\partial u' \partial v''}} \;
         \exp \left\{ \frac{\ri}{\hbar}S_{W} - \frac{1}{2}
          \bigl( |z''|^2 + |z'|^2\bigr) \right\} \,
\end{eqnarray}
with $S_W$ given by Eq.(\ref{glg83}) with $H_i$ replaced by $H_W$.

Of the three semiclassical approximations presented, the Weyl
approximation seems to be the most natural, since it involves the
classical hamiltonian directly and no corrections to the action.
However, this formula does not follow from the two most natural
forms of path integral proposed by Klauder and used in this section.
In the next section we propose a third form of path integral which
is constructed directly in terms of $H_W$ and whose semiclassical
limit is indeed the formula above. For a direct comparison between
these semiclassical formulas for short propagation times see
\cite{Pol03}.

\section{Coherent State Path Integrals with the Weyl Symbol}

The new form of path integral we describe in this section is based
on an expansion of the Hamiltonian in a continuous basis of
reflection operators $\hat{R}_x$ whose coefficients $H(x)$ are
exactly the Weyl symbol of $\hat{H}$. We first review the algebra of
reflection and translation operators in quantum mechanics
\cite{balazs}, following closely the presentation in
ref.\cite{Alf98}. We then use these results to construct the path
integral.

\subsection{Translation and Reflection Operators}

Consider the family of translation operators
\begin{equation}
\hat{T}_{\xi} = e^{\frac{i}{\hbar}(p\hat{q}-q\hat{p})} =
e^{ip\hat{q}/\hbar}\, e^{-iq\hat{p}/\hbar}\, e^{-iqp/2\hbar} =
e^{-iq\hat{p}/\hbar}\, e^{ip\hat{q}/\hbar}\, e^{+iqp/2\hbar}
\end{equation}
where $\xi=(q,p)$ is a point in phase space. It can be shown that
the $\hat{T}_{\xi}$ form a complete basis, in the sense that any
operator $\hat{A}$ can be expressed as
\begin{equation}
\hat{A} = \int \frac{{\rm d}\xi}{2\pi\hbar} A(\xi) \hat{T}_{\xi}.
\label{aex1}
\end{equation}
The Fourier transform of the operators $\hat{T}_{\xi}$ form a
complementary family of {\it reflection} operators $\hat{R}_x$ which
also form a basis:
\begin{equation}
\hat{R}_x = \frac{1}{4\pi\hbar} \int {\rm d}\xi \,
e^{\frac{i}{\hbar} (x \wedge \xi)} \, \hat{T}_{\xi}.
\end{equation}
where $x=(Q,P)$ and $x \wedge \xi = Pq-Qp$. In terms of these
operators we may write
\begin{equation}
\hat{A} = \int \frac{{\rm d}x}{\pi\hbar} A(x) \hat{R}_x =
\frac{1}{4\pi^2\hbar^2} \int {\rm d}\xi \,{\rm d}x A(x)\,
e^{\frac{i}{\hbar} (x \wedge \xi)}\, \hat{T}_{\xi}.
\label{aex2}
\end{equation}
When this expression is inverted to write that $A(x)$ in terms of
$\hat{A}$, we find precisely the Weyl representation, as given by
Eq.(\ref{wig1}). This is shown in Appendix A.

It is convenient to write some of these expressions in terms of
$\hat{a}$, $\hat{a}^\dagger$, $z$ and $z^*$ instead of $\hat{q}$,
$\hat{p}$, $q$ and $p$. We find that
\begin{equation}
\hat{T}_{\xi} = e^{(z\hat{a}^\dagger -z^*\hat{a})}\, =
e^{z\hat{a}^\dagger}\, e^{-z^*\hat{a}}\, e^{-|z|^2/2} \, ,
\end{equation}
which we recognize as the {\it displacement operator}
\cite{glauber,perelo,nieto} frequently used in quantum optics. Also
\begin{equation}
\langle z_k|\hat{T}_{\xi}|z_{k-1}\rangle = e^{z z^*_k - z^*z_{k-1} -
|z|^2/2} \, \langle z_k|z_{k-1}\rangle
\end{equation}
and
\begin{equation}
\langle z_k|\hat{A}|z_{k-1}\rangle = \frac{1}{4\pi^2\hbar^2} \int
{\rm d}x A(x) \langle z_k|z_{k-1}\rangle \int {\rm d} \xi \,
e^{\frac{i}{\hbar} (x \wedge \xi)} \, e^{z z^*_k - z^*z_{k-1} -
|z|^2/2}.
\end{equation}
Since the integral over $\xi=(q,p)$ is quadratic, it can be done
immediately. Defining
\begin{equation}
  w_k =  \frac{1}{\sqrt{2}}\left( \frac{Q}{b}+\ri
              \,\frac{P}{c} \right).
\end{equation}
(the index $k$ is added for later convenience) we find
\begin{equation}
\langle z_k|\hat{A}|z_{k-1}\rangle = 2\int \frac{{\rm d}w_k {\rm
d}w^*_k}{2\pi i} A(w_k,w^*_k) \, e^{-2|w_k|^2 + 2z^*_k w_k +
2z_{k-1}w^*_k - |z_k|^2/2 - |z_{k-1}|^2/2 - z^*_kz_{k-1}}
\label{zaz}
\end{equation}
where
\begin{equation}
\frac{{\rm d}w_k {\rm d}w^*_k}{2\pi i} = \frac{{\rm d}Q {\rm
d}P}{2\pi \hbar}.
\end{equation}

As the notation suggests, this expression will be our starting point
to construct the path integral. When $\hat{A}$ is replaced by the
infinitesimal propagator $e^{-i\hat{H}\tau/\hbar}\approx 1 -
i\hat{H}\tau/\hbar$ and a sequence of these matrix elements are
multiplied together, we will find that the all the $z_k$'s and
$z^*_k$'s appear only in quadratic forms and can be integrated over.
The resulting path integral will be written in the new variables
$w$.

\subsection{The Path Integral}

We start from
\begin{equation}
  K(z'',t;z',0) = \int \Bigl\{ \prod_{j=1}^{N-1}\frac{\rd z_j\rd z^*_j}{2\pi i}
                   \Bigr\} \prod_{j=1}^{N} \Bigl\{ \langle z_{j} |
                   \re^{-\frac{\ri }{\hbar}
                   \hat{H}(t_j)\tau} | z_{j-1} \rangle
                   \Bigr\}
\end{equation}
where $z_N=z"$, $z_0=z'$, $\tau$ is the time step, $N\tau=T$ and we
take $N$ to be even for convenience. The infinitesimal propagators
can be calculated with Eq.(\ref{zaz}) by simply replacing $A(x_k)$
by $e^{-iH(x_k)\tau/\hbar}$ where $H(x)$ is the Weyl symbol of
$\hat{H}$ calculated at $(Q_k,P_k)$. We obtain
\begin{equation}
\begin{array}{ll}
K&(z'',t;z',0) = 2^N \int \Bigl\{ \prod_{j=1}^{N}\frac{\rd w_j\rd
    w^*_j}{2\pi i} \Bigr\}
    \int \Bigl\{ \prod_{j=1}^{N-1}\frac{\rd z_j\rd z^*_j}{2\pi i}
    \Bigr\} \, \times \\ \\
&\exp{\{ \sum_{k=1}^{N} \left[-\frac{i}{\hbar}H_k\tau -2|w_k|^2 +
2z^*_k w_k + 2z_{k-1}w^*_k - \frac{|z_k|^2}{2} -
\frac{|z_{k-1}|^2}{2} - z^*_kz_{k-1} \right] \}}.
\end{array}
\end{equation}
where $H_k=H(w_k,w^*_k)$. The integrals over the $z_j$'s and the
$z^*_j$'s can be performed exactly. When this is done we find
\begin{equation}
\begin{array}{ll}
K(z'',t;z',0) =& \int \Bigl\{ \prod_{j=1}^{N}\frac{\rd w_j\rd
    w^*_j}{\pi i} \Bigr\} \,
    e^{ \phi_N - \frac{|z'|^2}{2} - \frac{|z''|^2}{2}} \\ \\
    &= \int {\cal D}[w,w^*] e^{\psi[w,w^*] + 2C[w,w^*]z''^*
    - 2C^*[w,w^*]z' - \frac{|z'|^2}{2} - \frac{|z''|^2}{2}+z'z''^*}
    \label{propw}
\end{array}
\end{equation}
where
\begin{equation}
\begin{array}{ll}
\phi_N  = &\sum_{k=1}^{N} \left[ -i\tau H_k/\hbar -2|w_k|^2 + 2z''^*
w_{N+1-k}(-1)^{k+1} + 2z' w^*_k(-1)^{k+1}  \right] \\ \\
&+4\sum_{k=1}^{N-1}\sum_{j=1}^{k}w^*_{k+1}w_{k+1-j}(-1)^{j+1} +
z'z''^*.
\end{array}
\label{actw}
\end{equation}
In the second line of (\ref{propw}) we have written the dependence
of the propagator on $z'$ and $z''^*$ explicitly and defined
\begin{equation}
\begin{array}{ll}
\psi_N  = &\sum_{k=1}^{N} \left[ -i\tau H_k/\hbar -2|w_k|^2 \right]
+4\sum_{k=1}^{N-1}\sum_{j=1}^{k}w^*_{k+1}w_{k+1-j}(-1)^{j+1}, \\
C_N &= \sum_{k=1}^{N} w_{N+1-k}(-1)^{k+1}.
\end{array}
\label{actpsi}
\end{equation}
%

\subsection{Alternative form and the limit of continuum}

Eqs.(\ref{propw}) and (\ref{actw}) correspond to the coherent state
path integral in the Weyl representation. It is very different from
the previous forms presented in section II in two respects: the
measure lacks a factor 2 in the denominator and, more importantly,
the exponent does not resemble an action at all. Although these
expressions appear to be the most practical for actual calculations,
we can manipulate the terms in $\phi_N$ to make it look more
familiar and similar to an action function. However, it is only when
we take the limit of the continuum that we really recognize the
action as part of the exponent. We shall do these manipulations now,
but we insist that Eqs.(\ref{propw}) and (\ref{actw}) are the direct
analogs of Eqs. (\ref{glg82a}) and (\ref{5mb8a}) for the Q and P
representations respectively. Although unusual, and perhaps more
complicated, we shall see that, in the semiclassical limit, the Weyl
form becomes the simplest of them all.

We show in Appendix B that the quadratic terms in $\phi_N$ can be
written as
\begin{equation}
\begin{array}{l}
-\sum_{k=1}^{N}2|w_k|^2 + 4 \sum_{k=1}^{N-1} \sum_{j=1}^{k}
w^*_{k+1}w_{k+1-j}(-1)^{j+1} \\
= 2 \sum_{k=1,3}^{N-1}\left[ w_k(w^*_{k+1}-w^*_k) -
w^*_{k+1}(w_{k+1}-w_k) \right] \\
 - 4\sum_{k=1,3}^{N-1}(w_{k+1}-w_k)
\sum_{l=k+1,k+3}^{N-2}(w^*_{l+2}-w^*_{l+1})
\end{array}
\label{alt1}
\end{equation}
where the sums on the right go in steps of two. The terms
proportional to $z'$ and $z''^*$ can also be re-written as
\begin{equation}
\begin{array}{l}
\sum_{k=1}^{N} w^*_k (-1)^{k+1} = -\sum_{k=1,3}^{N-1}
(w^*_{k+1}-w^*_k) \\
\sum_{k=1}^{N} w_{N+1-k} (-1)^{k+1} = \sum_{k=1,3}^{N-1}
(w_{k+1}-w_k).
\end{array}
\label{alt2}
\end{equation}
When these terms are replaced in the exponent we get
\begin{equation}
\begin{array}{ll}
\phi_N  = & 2 \sum_{k=1,3}^{N-1}\left[ w_k(w^*_{k+1}-w^*_k) -
w^*_{k+1}(w_{k+1}-w_k) \right] -\frac{i \tau}{\hbar} \sum_{k=1}^{N}
H_k  \\ &- 4\sum_{k=1,3}^{N-1}(w_{k+1}-w_k)
\sum_{l=k+1,k+3}^{N-2}(w^*_{l+2}-w^*_{l+1}) \\ & -
2z'\sum_{k=1,3}^{N-1} (w^*_{k+1}-w^*_k) + 2 z''^* \sum_{k=1,3}^{N-1}
(w_{k+1}-w_k) + z'z''^*.
\end{array}
\label{actalt}
\end{equation}
This is the alternative discrete version of $\phi_N$. Although not
much enlightening than the original form, Eq.(\ref{actw}), the first
line shows a closer resemblance to the usual action function. More
importantly, this expression is ready for the continuum limit.
Taking $N\rightarrow \infty$, $\tau\rightarrow 0$ with $N\tau = T$
we obtain
\begin{equation}
\begin{array}{ll}
\phi  = & -\frac{i \tau}{\hbar} \int_{0}^{T} H {\rm d}t +
\int_{0}^{T}\left( w \dot{w}^* - w^* \dot{w} \right){\rm d}t -
\int_{0}^{T}\dot{w}(t) \int_{t}^{T}\dot{w}^*(t'){\rm d}t' {\rm d}t \\
& - z'\int_{0}^{T} \dot{w}^*{\rm d}t + z''^* \int_{0}^{T} \dot{w}
{\rm d}t + z'z''^*.
\end{array}
\label{actalta}
\end{equation}
Notice that the factors of 2 and 4 compensate for the sums in steps
of two.

The integrals in the last term on the first line can be rewritten as
\begin{equation}
\int_{0}^{T} \dot{w}(t)[w^*(T)-w^*(t)]{\rm d}t = w^*(T)[w(T)-w(0)] -
\int_{0}^{T} \dot{w}(t) w^*(t) {\rm d}t.
\end{equation}
The last term above cancels one of the terms in Eq.(\ref{actalta}).
After performing the integrals on the second line of
Eq.(\ref{actalta}), making some simple rearrangements and an
integration by parts, we can write the exponent in the form
\begin{equation}
\begin{array}{ll}
\phi  = & \frac{i}{\hbar} S +
(z'-w(0))\left[w^*(0)+\frac{z''^*-w^*(T)}{2}\right] +
(z''^*-w^*(T))\left[w(T)+\frac{z'-w(0)}{2}\right]
\end{array}
\label{actaltb}
\end{equation}
where $S$ is the (complex) action \cite{Bar01}
\begin{equation}
S  =  \int_{0}^{T} \left[ \frac{i \hbar}{2} ( w^* \dot{w}  - w
\dot{w}^*) - H \right] {\rm d}t - \frac{i
\hbar}{2}(w^*(T)w(T)+w^*(0)w(0)). \label{actionf}
\end{equation}
Notice that the action in the coherent state representation is not
just the integral corresponding to $p\dot{q}-H$, but it includes
important boundary terms. Besides, the exponent $\phi$ of the path
integral is not just the action and also includes further boundary
terms. We shall see, however, that the extra terms in
Eq.(\ref{actaltb}) vanish in the semiclassical limit.

\section{Semiclassical Limit}

The semiclassical limit of the propagator is obtained by performing
the integrals over $w_k$ and $w^*_k$ with the stationary phase
approximation. Because the exponent $\phi_N$ is not a phase, but a
complex quantity, we use the terminology `stationary exponent
approximation'.

\subsection{The Stationary Exponent Condition}

Using Eq.(\ref{actw}) for $N$ even and $l\neq 1$ even we obtain
\begin{displaymath}
\frac{\partial \phi_N}{\partial w^*_l} = -\frac{i \tau}{\hbar}
\frac{\partial H_l}{\partial w^*_l} -2w_l -2z'
+4[w_{l-1}-w_{l-2}+\dots -w_2+w_1] \equiv 0
\end{displaymath}
and
\begin{displaymath}
\frac{\partial \phi_N}{\partial w^*_{l+1}} = -\frac{i \tau}{\hbar}
\frac{\partial H_{l+1}}{\partial w^*_{l+1}} -2w_{l+1} +2z'
+4[w_l-w_{l-1}+\dots +w_2-w_1]\equiv 0.
\end{displaymath}
Adding these two equations we obtain simply
\begin{equation}
-\frac{i}{\hbar} \frac{1}{2} \left[ \frac{\partial H_l}{\partial
w^*_l} +\frac{\partial H_{l+1}}{\partial w^*_{l+1}}\right] =
\frac{w_{l+1}-w_l}{\tau}. \label{eq1d}
\end{equation}
For $l=1$ we get
\begin{equation}
\frac{\partial \phi_N}{\partial w^*_1} = -\frac{i \tau}{\hbar}
\frac{\partial H_1}{\partial w^*_1} -2w_1 + 2z' \equiv 0.
\label{bc1d}
\end{equation}

For the derivatives with respect to $w_l$ we proceed in the same
way. For $l$ odd we get
\begin{displaymath}
\frac{\partial \phi_N}{\partial w_l} = -\frac{i \tau}{\hbar}
\frac{\partial H_l}{\partial w_l} -2w^*_l -2z''^*
+4[w^*_{l+1}-w^*_{l+2}+\dots -w^*_{N-1}+w^*_N] \equiv 0
\end{displaymath}
and
\begin{displaymath}
\frac{\partial \phi_N}{\partial w_{l+1}} = -\frac{i \tau}{\hbar}
\frac{\partial H_{l+1}}{\partial w_{l+1}} -2w^*_{l+1} +2z''^*
+4[w^*_{l+2}-w^*_{l+1}+\dots +w^*_{N-1}-w^*_N]\equiv 0.
\end{displaymath}
Adding the two equations we obtain
\begin{equation}
-\frac{i}{\hbar} \frac{1}{2} \left[ \frac{\partial H_l}{\partial
w_l} +\frac{\partial H_{l+1}}{\partial w_{l+1}}\right] =
-\frac{w^*_{l+1}-w^*_l}{\tau}. \label{eq2d}
\end{equation}
Finally for $l=N$ we get
\begin{equation}
\frac{\partial \phi_N}{\partial w_N} = -\frac{i \tau}{\hbar}
\frac{\partial H_1}{\partial w_N} -2w^*_N + 2z''^* \equiv 0.
\label{bc2d}
\end{equation}

Taking the continuum limit and using the $u$ and $v$ variables in
the place of $w$ and $w^*$, Eqs.(\ref{eq1d}), (\ref{eq2d}),
(\ref{bc1d}) and (\ref{bc2d}) become
\begin{eqnarray}
  \label{eqmov}
 \ri \hbar\dot{u}&= + \displaystyle{\frac{\partial H_W}{\partial
 v}} ~, \qquad
 \ri \hbar\dot{v}&= - \displaystyle{\frac{\partial H_W}{\partial u}}
\end{eqnarray}
with boundary conditions
\begin{equation}
\label{bc} u(0) = z'~, \qquad v(T) = {z''}^\star \;.
\end{equation}
The average of the derivatives at consecutive time steps that
appears in the left side of equations (\ref{eq1d}) and (\ref{eq2d})
resemble the stationary conditions obtained in \cite{coelho}. In
that case, however, one of the derivatives involved $H_P$ and the
other $H_Q$.

\subsection{Expansion Around the Stationary Trajectory}

Let $w_k^0$ and $w^{*0}_k$ represent the stationary trajectory and
$w_k^0+\xi_k$ and $w^{*0}_k+\xi^*_k$ a nearby path. Expanding the
exponent up to second order around the stationary trajectory we get
\begin{equation}
\phi_N = \phi_N^0 + \delta^2 \phi_N + O(3)
\end{equation}
(the first order term is zero) with
\begin{equation}
\begin{array}{ll}
\delta^2 \phi_N &=
\sum_{k=1}^N\left\{-\frac{i\tau}{2\hbar}[A_k\xi_k^2 +
2C_k\xi_k\xi^*_k + B_k\xi^{*2}_k] - 2\xi_k\xi^*_k\right\}
+4\sum_{k=1}^{N-1}\xi^*_{k+1}\sum_{j=1}^{k}\xi_{k+1-j}(-1)^{j+1} \\
\\ &\equiv -\frac{1}{2} X^T \tilde{\Delta}_N X \label{quad}
\end{array}
\end{equation}
where $X^T = (\xi_N,\xi^*_N,\xi_{N-1},\dots,\xi_1,\xi_1^*)$ and
\begin{equation}
\begin{array}{ll}
A_k = \frac{\partial^2 H_k}{\partial w_k^2}~, \quad B_k =
\frac{\partial^2 H_k}{\partial w_k^{*2}}~, \quad C_k =
\frac{\partial^2 H_k}{\partial w_k \partial w_k^*}
\end{array}
\end{equation}
are calculated at the stationary trajectory.

When the limit of the continuum is taken, the boundary conditions
Eq.(\ref{bc}) kill the extra terms in the exponent $\phi$,
Eq.(\ref{actaltb}), which becomes simply the action of the complex
trajectory. Therefore the semiclassical propagator becomes
\begin{equation}
K_W(z',z'',T) = e^{\frac{i}{\hbar}S_W -\frac{1}{2}(|z'|^2 +
|z''|^2)} \, \lim_{N\rightarrow \infty} \frac{2^N}{\sqrt{(-1)^N {\rm
det}(\tilde{\Delta}_N)}}. \label{det1}
\end{equation}
As usual, the calculation of the determinant of the quadratic form
is the most lengthy step of the semiclassical calculation. In this
case the calculation is particularly tricky, because of the double
sum in the last term of the first line of equation (\ref{quad}). To
avoid losing the focus with this lengthy algebra here we do the
calculation in the Appendix C. The final result is indeed the
conjectured formula, Eq.(\ref{glg119b}) that we repeat here:
\begin{eqnarray}
  K_W(z'',t;z',0) =
         \sqrt{\frac{\ri}{\hbar}\frac{\partial^2 S_{W}}
         {\partial u' \partial v''}} \;
         \exp \left\{ \frac{\ri}{\hbar}S_{W} - \frac{1}{2}
          \bigl( |z''|^2 + |z'|^2\bigr) \right\} \,. \label{kwsemi}
\end{eqnarray}
Of course, if there is more than one stationary trajectory, one
should sum over all the contributing ones.

\section{A Comparison Between the Three Forms of Path Integral}

In principle, all discrete forms of path integrals given by
Eqs.(\ref{glg82a}), (\ref{5mb8a}) and (\ref{propw}) are quantum
mechanically equivalent. For fixed $N$, however, they are not
identical and in the limit $N\rightarrow\infty$ there are well known
convergence problems, making the comparison difficult. In order to
illustrate the differences between the three forms we shall study
the discrete propagators for the simple harmonic oscillator. The
Hamiltonian operator is
\begin{displaymath}
\hat{H} = -\frac{\hbar^2}{2m} \frac{\partial^2}{\partial x^2} +
\frac{m\omega^2 x^2}{2} = \hbar\omega\left(a^{\dagger} a +
\frac{1}{2}\right)
\end{displaymath}
and, choosing the coherent state width as $b=\sqrt{\hbar/m\omega}$,
the classical symbols, in the $u$ and $v$ variables, are
\begin{displaymath}
H_Q = \hbar\omega\left(u v + \frac{1}{2}\right) \qquad H_P =
\hbar\omega\left(u v - \frac{1}{2}\right) \qquad H_W = \hbar\omega
uv.
\end{displaymath}

Using $H_W$ in the stationary conditions (\ref{eq1d})-(\ref{bc2d})
we obtain the stationary path
\begin{displaymath}
w_k = \frac{{\alpha^*}^{k-1}}{\alpha^k}\, z' \qquad \quad w^*_k =
\frac{{\alpha^*}^{N-k}}{\alpha^{N-k+1}}\, z''^*
\end{displaymath}
where
\begin{displaymath}
\alpha \equiv 1+i\tau\omega/2.
\end{displaymath}
The calculation of the phase $\phi_N^0$ at the stationary trajectory
is lengthy but involves only simple geometric sums. Several
simplifications occur when all the terms in $\phi_N^0$ are added
together and the result is
\begin{displaymath}
\phi_N^0 = \left(\frac{\alpha^*}{\alpha}\right)^N z' z''^*
-\frac{1}{2}\left(|z'|^2+|z''|^2\right).
\end{displaymath}
The determinant of the quadratic form is calculated in Appendix C
and results in (see Eqs.(\ref{deltaa}) and (\ref{detosc}))
\begin{equation}
\det \tilde{\Delta}_N = 2^{2N} i^{2N} \alpha^{2N}.
\end{equation}

Putting everything together we obtain
\begin{equation}
K_W(z'',z',T) = (1+i\tau\omega/2)^{-N} \,
e^{\left(\frac{1-i\tau\omega/2}{1+i\tau\omega/2}\right)^N z'z''^* -
|z'|^2/2 -|z''|^2/2 }
\end{equation}
which clearly converges to the exact propagator as $\tau \rightarrow
0$. Doing similar calculations for the Q and P propagators we find
\begin{equation}
K_Q(z'',z',T) = e^{-i\omega T/2 +\left(1-i\tau\omega\right)^N
z'z''^* - |z'|^2/2 -|z''|^2/2 }
\end{equation}
and
\begin{equation}
K_P(z'',z',T) = (1+i\tau\omega)^{-N} \, e^{i\omega T/2 +
\frac{z'z''^*}{(1+i\tau\omega)^N}  - |z'|^2/2 - |z''|^2/2 }
\end{equation}
which also converge to the exact result. Notice that the overall
phase $-i\omega T/2$ comes out exact for $K_Q$ even in the discrete
form. However, the term multiplying $z'z''^*$, which goes to
$e^{-i\omega T}$ as $N\rightarrow \infty$, converges much faster for
$K_W$ then the corresponding terms in $K_Q$ or $K_P$. Moreover, for
any finite value of $N$, this term has unit modulus in $K_W$, while
its modulus is larger than one for $K_Q$ and smaller than one in
$K_P$. Just for the sake of comparison let us call $\mu$ this
coefficient. Taking $\omega T =2\pi$ and $N=100$ we find $\mu_Q
\approx 1.22 + 0.01i$, $\mu_P \approx 0.82 + 0.007i$ and $ \mu_W
\approx 0.999998 + 0.002i$. This suggests that the new path integral
representation should be better than the two KS forms for numerical
evaluations.

\section{Connecting the Wigner and the Husimi propagators}

In this section we show that the Weyl representation of the
evolution operator
\begin{equation}
U(q,p,T) = \int \langle q-s/2|e^{-i\hat{H}T/\hbar}|q+s/2\rangle
\,e^{ips/\hbar} {\rm d} s \label{wig1a}
\end{equation}
can be directly related to the path integral representation derived
in the section III. This is an interesting formal result that was
also obtained by Ozorio de Almeida in section 6 of \cite{Alf98}
starting from the opposite direction, i.e., from the path integral
representation of $U$. The result provides an explicit connection
between these two famous phase space representations of quantum
mechanics. As we shall see, the connection is very simple when
written in terms of path integrals.

We start by re-writing Eq.(\ref{wig1a}) as
\begin{equation}
\begin{array}{ll}
U(q,p,T) &= \int \langle q-s/2|z''\rangle\langle z''|\hat{U}
|z'\rangle \langle z'|q+s/2\rangle \,e^{ips/\hbar} {\rm d} s
\frac{\rd z'\rd z'^*}{2\pi i} \frac{\rd z''\rd z''^*}{2\pi i} \\ \\
&=\int {\cal D}[w,w^*] e^{\psi} \int {\rm d}s e^{ips/\hbar} \int
\frac{\rd z'\rd z'^*}{2\pi i} \frac{\rd z''\rd z''^*}{2\pi i}
\langle \beta|z''\rangle \langle z'|\alpha \rangle \, \times
\\  &\qquad \qquad \exp{\left[-\frac{|z'|^2}{2} -\frac{|z''|^2}{2} +z'z''^*
+ 2Cz''^* - 2C^*z'\right]}\label{wig2}
\end{array}
\end{equation}
where we used Eqs.(\ref{propw}) and (\ref{actpsi}) and defined
$\alpha = q+s/2$ and $\beta=q-s/2$ in the second line. The integrals
in $z'$ and $z''$ are quadratic and be performed analytically. The
integral over $z'$ is straightforward and gives
\begin{equation}
\begin{array}{ll}
U(q,p,T) &= \frac{1}{\pi^{1/4}b^{1/2}} \int {\cal D}[w,w^*] e^{\psi}
\int {\rm d}s e^{ips/\hbar} \int \frac{\rd z''\rd z''^*}{2\pi i}
 \, \times \\
& \qquad \qquad \langle \beta|z''\rangle \,
\exp{\left[-\frac{|z''|^2}{2} \,+\, 2Cz''^* -\frac{\alpha^2}{2b^2}
-\frac{z''^*-2C^*}{2} + \frac{\alpha\sqrt{2}}{b}(z''^*-2C^*)
\right]}. \label{wig3}
\end{array}
\end{equation}
It can be seen by inspection that the exponent in the second line
above can be written as
\begin{equation}
\pi^{1/4}b^{1/2} \, \langle z''|\alpha + A\rangle e^{-B}
\label{wig4}
\end{equation}
with $A=b\sqrt{2}(C+C^*)$ and $B=-A^2/2b^2-A\alpha/b^2+2C^{*2} +
2\sqrt{2}\alpha C^*/b$. When (\ref{wig4}) is substituted into
(\ref{wig3}) the integral in $z''$ produces $\langle
\beta|\alpha+A\rangle = \delta(\alpha -\beta +A) = \delta(s+A)$. The
delta function takes care of the integral over $s$ and after some
simplifications we obtain simply
\begin{equation}
U(q,p,T) = \int {\cal D}[w,w^*] e^{\psi + 2Cz_x^* - 2C^*z_x + 2
|C|^2} \label{wig5}
\end{equation}
where $z_x=(q/b+ipb/\hbar)/\sqrt{2}$. Comparing with
Eq.(\ref{propw}) shows that the path integral for $U(q,p,T)$ is
directly related to that for $K(z_x,z_x,T)$. Indeed, the path
integral for $U$ has a single extra term $2|C|^2$ with respect to
the $K$. One might say that this terms promotes the `unsmoothing' of
the coherent state propagator. Conversely, the diagonal coherent
state propagator has the extra term $-2|C|^2$ with respect to the
$U$, smoothing it out. This result was also obtained by Ozorio de
Almeida in section 6 of \cite{Alf98}. The coefficient $C$ can
actually be interpreted as the Wigner chord linking the ends of a
polygon in phase space whose sides are given by the $Q_k$ and $P_k$
variables in $w_k$. We can also calculate explicitly the two terms
that involve $q$ and $p$ in (\ref{wig5}). Using the definition of
$C$ in Eq.(\ref{actpsi}) we find that
\begin{equation}
2Cz_x^* - 2C^*z_x = \sum_{k=1}^N \frac{2i}{\hbar} (Q_k p-P_k q)
\label{wig7}
\end{equation}
which is the sum of the symplectic areas between $X_k=(Q_k,P_k)$ and
$x=(q,p)$ and is independent of the width $b$.

\begin{appendix}
\section{Expansion in Reflection and Translation Operators}

This appendix follows closely the demonstration in \cite{Alf98}. A
comparison between Eqs.(\ref{aex1}) and (\ref{aex2}) shows that
\begin{equation}
A(\xi) = \frac{1}{2\pi\hbar}\int {\rm d}x A(x) e^{\frac{i}{\hbar}
x\wedge \xi}
\end{equation}
and, inverting the Fourier transform,
\begin{equation}
A(x) = \frac{1}{2\pi\hbar}\int {\rm d}\xi A(\xi) e^{-\frac{i}{\hbar}
x\wedge \xi} \label{xxi} \,.
\end{equation}
Using Eq.(\ref{aex1}) again in the coordinate representation we
obtain
\begin{equation}
\begin{array}{ll}
\langle q_+|\hat{A}|q_-\rangle &= \int \frac{{\rm d}\xi}{2\pi\hbar}
A(\xi) \langle q_+|\hat{T}_{\xi}|q_-\rangle \\
&=\int \frac{{\rm d}q{\rm d}p}{2\pi\hbar} A(q,p) \delta(q_+-q_--q)\,
e^{\frac{i}{\hbar} p \left(q_-+\frac{q}{2}\right)} \\
&=\int \frac{{\rm d}p}{2\pi\hbar} A(p,q_+-q_-) \,
e^{\frac{i}{\hbar}\frac{q_++q_-}{2}p}
\end{array}
\end{equation}
This Fourier transform can be inverted as follows: we define
$q'=q_+-q_-$, $\bar{Q}=(q_++q_-)/2$, multiply both sides by
$e^{-ip'\bar{Q}/\hbar}$ and integrate over $\bar{Q}$. The integral
over $\bar{Q}$ on the right-hand-side yields a delta function on
$p-p'$ and we obtain
\begin{equation}
A(\xi) = \int {\rm d}\bar{Q} \langle \bar{Q} + q/2 |\hat{A}|\bar{Q}
- q/2\rangle e^{-\frac{i}{\hbar} p\bar{Q}}  \,.
\end{equation}
where $(q',p')$ has been changed back to $(q,p)$. Finally we use
Eq.(\ref{xxi}) to get $A(x)$:
\begin{equation}
\begin{array}{ll}
A(x) &= \frac{1}{2\pi\hbar} \int {\rm d}q {\rm d}p {\rm d}\bar{Q} \,
e^{\frac{i}{\hbar}p(Q-\bar{Q})-\frac{i}{\hbar}Pq} \, \langle \bar{Q}
+ q/2 |\hat{A}|\bar{Q} - q/2\rangle \\
& = \int {\rm d}q \, e^{-\frac{i}{\hbar}Pq} \, \langle Q + q/2
|\hat{A}|Q - q/2\rangle
\end{array}
\end{equation}
which is the same as Eq.(\ref{wig1}).

\section{Proof of Eq.(\ref{alt1})}

First we re-write, for $N$ even,
\begin{equation}
\begin{array}{ll}
4 \sum_{k=1}^{N-1} \sum_{j=1}^{k} & w^*_{k+1}w_{k+1-j}(-1)^{j+1} =
\\ & 4w_2^*w_1 \, + \\ & 4w_3^*[w_2-w_1] \, +\\ &4w_4^*[w_3-(w_2-w_1)] \,
+ \\ & 4w_5^*[(w_4-w_3)+(w_2-w_1)] \,+ \\
&4w_6^*[w_5-(w_4-w_3)-(w_2-w_1)] + \\ & \qquad \vdots  \\ &
4w_N^*[w_{N-1}-(w_{N-2}-w_{N-3})-\dots -
(w_2-w_1)] \, = \\ \\
&4[w_2^*w_1+w_4^*w_3+w_6^*w_5 + \dots + w_N^*w_{N-1}] \,- \\ &
4(w_2-w_1)[(w_4^*-w_3^*)+(w_6^*-w_5^*)+ \dots +(w_N^*-w_{N-1}^*)]
\,- \\ & 4(w_4-w_3)[(w_6^*-w_5^*)+(w_8^*-w_7^*)+ \dots
+(w_N^*-w_{N-1}^*)] \, - \\ & \qquad \vdots  \\
&4(w_{N-2}-w_{N-3})[w_N^*-w_{N-1}^*] \, = \\ \\
&4\sum_{k=1,3}^{N-1}w_{k+1}^*w_k - 4\sum_{k=1,3}^{N-3}(w_{k+1}-w_k)
\sum_{l=k+1,k+3}^{N-2}(w^*_{l+2}-w^*_{l+1})
\end{array}
\end{equation}
The second term is already in the form needed for Eq.(\ref{alt1}).
The first term is now modified as follows: half of it remains
unchanged and, in the second half, we add and subtract terms as in
\begin{equation}
w^*_{k+1}w_k = w^*_{k+2}w_{k+1} -[w_{k+1}(w^*_{k+2}-w^*_{k+1}) +
w^*_{k+1}(w_{k+1}-w^*_k)]
\end{equation}
for $k=1,3,\dots,N-3$ only. We obtain
\begin{equation}
\begin{array}{ll}
4\sum_{k=1,3}^{N-1}w_{k+1}^*w_k  &= 2\sum_{k=1,3}^{N-1}w_{k+1}^*w_k
+ 2\sum_{k=1,3}^{N-3}w_{k+2}^*w_{k+1} - \\
& 2\sum_{k=1,3}^{N-3}[w_{k+1}(w^*_{k+2}-w^*_{k+1}) +
w^*_{k+1}(w_{k+1}-w^*_k)] + 2w^*_Nw_{N-1}
\end{array}
\end{equation}
We finally add $-2\sum_{k=1}^N w^*_k w_k$. The part of this sum
containing odd k's goes together with the first sum above. The even
k's up to $N-2$ goes with the second sum. We get
\begin{equation}
\begin{array}{l}
4\sum_{k=1,3}^{N-1}w_{k+1}^*w_k -2\sum_{k=1}^N w^*_k w_k = \\
2\sum_{k=1,3}^{N-1}w_k(w_{k+1}^*-w^*_k)
+ 2\sum_{k=1,3}^{N-3}w_{k+1}(w_{k+2}^*-w_{k+1}^*) - \\
2\sum_{k=1,3}^{N-3}[w_{k+1}(w^*_{k+2}-w^*_{k+1}) +
w^*_{k+1}(w_{k+1}-w_k)] - 2w^*_N (w_N-w_{N-1})
\end{array}
\end{equation}
The second term in the second line cancels against the first term of
the third line. After incorporating the last term into the sum we
get
\begin{equation}
\begin{array}{l}
4\sum_{k=1,3}^{N-1}w_{k+1}^*w_k -2\sum_{k=1}^N w^*_k w_k =
2\sum_{k=1,3}^{N-1}[w_k(w_{k+1}^*-w_k^*)- w^*_{k+1}(w_{k+1}-w_k)]
\end{array}
\end{equation}
%

\section{Calculation of the Determinant}

The quadratic form in Eq.(\ref{det1}) is defined by the matrix
\begin{displaymath}
    \begin{pmatrix}
      i \tau A_N/\hbar & i \tau C_N/\hbar + 2& 0 & 0 & 0 & 0 & 0 &\dots \\
      i \tau C_N/\hbar + 2 & i\tau B_N/\hbar & -4 & 0 & 4 & 0 & -4 &\dots \\
      0 & -4 & i \tau A_{N-1}/\hbar  & i \tau C_{N-1}/\hbar + 2 & 0 & 0 & 0 &\dots \\
      0 & 0 &i \tau C_{N-1}/\hbar + 2 & i\tau B_{N-1}\hbar & -4 & 0 & 4 &\dots \\
      0 & 4 & 0 & -4 &   &   &  & \dots \\
      0 & 0 & 0 & 0 &   &   &  & \dots \\
      \vdots & \vdots & \vdots & \vdots & \vdots & \vdots &  &\\
        &   &   &   &   &   & 0 & 0 \\
        &   &   &   &   &   & -4 & 0 \\
        &   &   &   & \dots  & -4  & i \tau A_1/\hbar & i \tau C_1/\hbar + 2 \\
        &   &   &   & \dots  &  0 & i \tau C_1/\hbar + 2 & i \tau B_1/\hbar \\
    \end{pmatrix}
\end{displaymath}
whose determinant, $\det \tilde{\Delta}_N$, we seek. To simplify the
notation we will drop the $\det$ symbol in this appendix and use
simply $\tilde{\Delta}_N$ for $\det \tilde{\Delta}_N$. It is useful
to factor $2i$ out of each element and call the new determinant
$\Delta_N$. Of course
\begin{equation}
\tilde{\Delta}_N = 2^{2N} i^{2N} \Delta_N. \label{deltaa}
\end{equation}
This cancels both the $2^N$ and the sign $(-1)^N$ in
Eq.(\ref{det1}), leaving only $\Delta_N$. Next we do the following
sequence of operations that do not change the value of the determinant:\\
\noindent column 2 $\rightarrow$ column 2 + column 4 \\
\noindent column 4 $\rightarrow$ column 4 + column 6 \\
\noindent \vdots \\
\noindent column N-2 $\rightarrow$ column N-2 + column N-4 \\
\noindent line 2 $\rightarrow$ line 2 + line 4 \\
\noindent line 4 $\rightarrow$ line 4 + line 6 \\
\noindent \vdots \\
\noindent line N-2 $\rightarrow$ line N-2 + line N-4. \\
This put the matrix in block tri-diagonal form:
\begin{displaymath}
    \begin{pmatrix}
      \frac{\tau A_N}{2\hbar} & \frac{\tau C_N}{2\hbar} - i& 0 & 0 & 0 & 0
                     & 0 &\dots \\
      \frac{\tau C_N}{2\hbar} - i & \frac{\tau(B_N+B_{N-1})}{2\hbar}
                     & \frac{\tau C_{N-1}}{2\hbar} + i
                     & \frac{\tau B_{N-1}}{2\hbar} & 0 & 0 & 0 &\dots \\
      0 & \frac{\tau C_{N-1}}{2\hbar} + i & \frac{\tau A_{N-1}}{2\hbar}
                     & \frac{\tau C_{N-1}}{2\hbar} - i
                     & 0 & 0 & 0 &\dots \\
      0 & \frac{\tau B_{N-1}}{2\hbar} & \frac{\tau C_{N-1}}{2\hbar} - i
                     & \frac{\tau(B_{N-1}+B_{N-2})}{2\hbar}
                     & \frac{\tau C_{N-2}}{2\hbar} + i
                     & \frac{\tau B_{N-2}}{2\hbar} & 0 &\dots \\
      0 & 0 & 0 & \frac{\tau C_{N-2}}{2\hbar} + i &   &   &  & \dots \\
      0 & 0 & 0 & \frac{\tau B_{N-2}}{2\hbar} &   &   &  & \dots \\
      \vdots & \vdots & \vdots & \vdots & \vdots & \vdots &  &\\
        &   &   &   &   &   & 0 & 0 \\
        &   &   &   &   &   & \frac{\tau C_1}{2\hbar} + i & \frac{\tau B_1}{2\hbar} \\
        &   &   &   & \dots  & \frac{\tau C_1}{2\hbar} + i  & \frac{\tau A_1}{2\hbar}
                     & \frac{\tau C_1}{2\hbar} - i \\
        &   &   &   & \dots  &  \frac{\tau B_1}{2\hbar} & \frac{\tau C_1}{2\hbar} - i
                     & \frac{\tau B_1}{2\hbar} \\
    \end{pmatrix}
\end{displaymath}

We can now compute the determinant using Laplace's method. Let
$\Gamma_N$ be the determinant obtained from the matrix above by
removing the first line and the first column. The two determinants
$\Delta_N$ and $\Gamma_N$ satisfy the following recursion relation:
\begin{equation}
\begin{array}{ll}
\Delta_N &= \frac{\tau A_N}{2\hbar} \Gamma_N - \left(\frac{\tau
C_N}{2\hbar} - i\right)^2 \Delta_{N-1} \\
\Gamma_N &= \frac{\tau(B_N+B_{N-1})}{2\hbar} \Delta_{N-1} -
\left(\frac{\tau C_{N-1}}{2\hbar} + i\right)^2 \Gamma_{N-1} + \\
&\left(\frac{\tau^2 C^2_{N-1}}{4\hbar^2} + 1\right)\frac{\tau
B_{N-1}}{2\hbar} \Delta_{N-2} + \frac{\tau B_{N-1}}{2\hbar}
\left[1+\frac{\tau^2}{4\hbar^2}(C_{N-1}^2-A_{N-1}B_{N-1})\right]
\Delta_{N-2} \label{appc1}
\end{array}
\end{equation}

Keeping only terms of first order in $\tau$ and taking the limit
$\tau \rightarrow 0$ we find
\begin{equation}
\begin{array}{ll}
\frac{\Delta_N -\Delta_{N-1}}{\tau}&= \frac{A_N}{2\hbar} \Gamma_N +
i\frac{C_N}{\hbar}\Delta_{N-1} + {\cal O}(\tau^2) \\
\frac{\Gamma_N -\Gamma_{N-1}}{\tau}&= \frac{(B_N+B_{N-1})}{2\hbar}
\Delta_{N-1} -i\frac{C_{N-1}}{\hbar} \Gamma_{N-1} +
\frac{B_{N-1}}{\hbar} \Delta_{N-2} + {\cal O}(\tau^2)
\end{array}
\end{equation}
or
\begin{equation}
\begin{array}{ll}
\dot{\Delta} &= \frac{A}{2\hbar} \Gamma +
i\frac{C}{\hbar}\Delta \\
\dot{\Gamma}&= \frac{2B}{\hbar} \Delta -i\frac{C}{\hbar} \Gamma
\label{appc2}
\end{array}
\end{equation}
with initial conditions $\Delta(0)=1$ and $\Gamma(0)=0$.

Notice that in the case of the harmonic oscillator $H_k = \hbar
\omega w_k w^*_k$ and, therefore, $A_k=B_k=0$ and $C_k=\hbar
\omega$. In this case Eqs.(\ref{appc1}) can be solved exactly,
without the need to take the continuum limit. We find simply
\begin{equation}
\Delta_N = - \left(\frac{\tau C_N}{2\hbar} - i\right)^2 \Delta_{N-1}
= \left(1+\frac{i\omega\tau}{2}\right)^2 \Delta_{N-1}
\end{equation}
which can be iterated to give
\begin{equation}
\Delta_N = \left(1+\frac{i\omega\tau}{2}\right)^{2N}. \label{detosc}
\end{equation}

To solve Eqs.(\ref{appc2}) in the general case we need a last change
of variables $\Omega \equiv 2i \Delta$. In the new variable we get
\begin{equation}
\begin{array}{ll}
\dot{\Omega} &= i\frac{A}{\hbar} \Gamma +
i\frac{C}{\hbar}\Omega \\
\dot{\Gamma}&= -i\frac{B}{\hbar} \Omega -i\frac{C}{\hbar} \Gamma
\label{appc3}
\end{array}
\end{equation}
with $\Omega(0)=2i$ and $\Gamma(0)=0$. Identifying $\Gamma$ with $u$
and $\Omega$ with $v$, we recognize these equations immediately as
the equations of motion (\ref{eqmov}) linearized around the
stationary trajectory. The solution we seek,
$\Delta(T)=\Omega(T)/2i$ can be obtained with the help of the
relations
\begin{equation}
-i\hbar u'' = \frac{\partial S}{\partial v''} \qquad -i\hbar v' =
\frac{\partial S}{\partial u'}
\end{equation}
where we use a single prime for quantities calculated at $t=0$ and a
double prime when $t=T$. A variation in the second of these
equations leads to
\begin{equation}
-i\hbar \delta v' = \frac{\partial^2 S}{\partial u'^2} \delta u' +
\frac{\partial^2 S}{\partial u' \partial v''} \delta v'' .
\end{equation}
Using $\delta u' = \Gamma(0) = 0$, $\delta v'' = \Omega(T)$ and
$\delta v' = \Omega(0)=2i$ we get
\begin{equation}
\Omega(T) = 2i (-i\hbar) \left(\frac{\partial^2 S}{\partial u'
\partial v''}\right)^{-1}
\end{equation}
and
\begin{equation}
\Delta = \left(\frac{i}{\hbar} \frac{\partial^2 S}{\partial u'
\partial v''}\right)^{-1}.
\end{equation}

\end{appendix}

\vspace{1cm}
\noindent ACKNOWLEDGMENTS

It is a pleasure to thank Alfredo Ozorio de Almeida for suggestions
and a careful reading of this manuscript. MAMA acknowledges
financial support from the Brazilian agencies FAPESP and CNPq.



\begin{thebibliography}{99}

\bibitem{Klau85}
J.~R. Klauder and B.~S. Skagerstam,  {\em Coherent States,
Applications in Physics and Mathematical Physics},   World
Scientific, Singapore, 1985.

\bibitem{Bar01} M. Baranger, M. A. M. de Aguiar, F. Keck, H. J.
Korsch and B. Schellhaaß, J. Phys. A {\bf 34} (2001) 7227.

\bibitem{coelho} L.C. dos Santos and M. A. M. de Aguiar,
Braz. J. Phys. {\bf 35} (2005) 175.

\bibitem{balazs} N.L. Balazs and B.K. Jennings, Phys. Rep. {\bf 104}
(1984) 347.

\bibitem{Alf98} A.M. Ozorio de Almeida, Phys. Rep. {\bf 295}
(1998) 265.

\bibitem{grosche} C. Grosche and F. Steiner, {\em Handbook of Feynman
Path Integrals (Springer Tracts in Modern Physics)}, Springer-Verlag
Telos (1998).

\bibitem{Hill84}
M.~Hillery, R.~F. O`Connel, M.~O. Scully, and E.~P. Wigner,{\em
Phys. Rep.}  {\bf 106}  (1984)  121.

\bibitem{Klau78}
J.~R. Klauder,  {\it Continuous Representations and Path Integrals,
Revisited}, in G.~J. {Papa\-do\-pou\-los} and J.~T. Devreese,
editors, {\em Path Integrals}, NATO Advanced Study Institute, Series
B: Physics, page~5, New York, 1978. Plenum.

\bibitem{Klau79}
J.~R. Klauder, {\em Phys. Rev. D}  {\bf 19(8)}  (1979)   2349.

\bibitem{Klau87a}
J.~R. Klauder,  {\it Some Recent Results on Wave Equations, Path
Integrals and Semiclassical Approximations},  in G.~Papanicolaou,
editor, {\em Random Media}, Random Media. Springer, 1987.

\bibitem{Weis82b}
Y.~Weissman,  {\em J. Chem. Phys.}  {\bf 76}  (1982) 4067.

\bibitem{Ada89} S. Adachi, Ann. of Phys., {\bf 195} 45 (1989).

\bibitem{Rib04} A.D. Ribeiro, M.A.M. de Aguiar and M. Baranger,
Phys. Rev. E {\bf 69} (2004) 66204.

\bibitem{Pol03} E. Pollak and J. Shao, J. Phys. Chem. A {\bf 107}
(2003) 7112.

\bibitem{glauber} R.J. Glauber, Phys. Rev. {\bf 131} (1963) 2766.

\bibitem{perelo} A. M. Perelomov, Comm. Math. Phys. {\bf 26} (1972)
222.

\bibitem{nieto} M. M. Nieto and L.M. Simmons, Jr.,  Phys. Rev. D
{\bf 20} (1979) 1321.

\end{thebibliography}
\end{document}